\documentclass{article}

\usepackage{spconf,amsmath,graphicx}
\usepackage{algorithm,algorithmicx,algpseudocode}
\usepackage{amsmath,amssymb}
\usepackage{multirow}
\usepackage{graphicx}
\usepackage{epstopdf}
\usepackage{hyperref}
\usepackage{color,soul}
\usepackage{tabularx}
\usepackage{cite}
\usepackage{amsmath}
\usepackage{subcaption}
\usepackage[english]{babel}
\usepackage{algorithm,algorithmicx,algpseudocode}
\usepackage{enumitem}
\usepackage{breqn}
\usepackage{xcolor}
\usepackage{makecell}
\usepackage{eqparbox}
\usepackage{color}
\usepackage{booktabs}
\usepackage{multirow}
\DeclareMathOperator{\E}{\mathbb{E}}
\newdimen{\algindent}

\title{Speaker diarization using latent space clustering in generative adversarial network}
\name{\begin{tabular}{c}Monisankha Pal$^{1}$, Manoj Kumar$^{1}$, Raghuveer Peri$^{1}$, Tae Jin Park$^{1}$ So Hyun Kim$^{2}$, Catherine Lord$^{3}$, \\Somer Bishop$^{4}$, Shrikanth Narayanan$^{1}$\end{tabular}}

\address{
  $^{1}$Signal Analysis and Interpretation Laboratory, University of Southern California\\
  $^{2}$Center for Autism and the Developing Brain, Weill Cornell Medicine\\
  $^{3}$Semel Institute of Neuroscience and Human Behavior, University of California Los Angeles\\
  $^{4}$Department of Psychiatry, University of California, San Francisco
}

\begin{document}
\ninept

\linespread{0.9}
\maketitle
\begin{abstract}
In this work, we propose deep latent space clustering for speaker diarization using generative adversarial network (GAN) back-projection with the help of an encoder network. The proposed diarization system is trained jointly with GAN loss, latent variable recovery loss, and a clustering-specific loss. It uses x-vector speaker embeddings at the input, while the latent variables are sampled from a combination of continuous random variables and discrete one-hot encoded variables using the original speaker labels. We benchmark our proposed system on the AMI meeting corpus, and two child-clinician interaction corpora (ADOS and BOSCC) from the autism diagnosis domain. ADOS and BOSCC contain diagnostic and treatment outcome sessions respectively obtained in clinical settings for verbal children and adolescents with autism. Experimental results show that our proposed system significantly outperform the state-of-the-art x-vector based diarization system on these databases. Further, we perform embedding fusion with x-vectors to achieve a relative DER improvement of 31\%, 36\% and 49\% on AMI eval, ADOS and BOSCC corpora respectively, when compared to the x-vector baseline using oracle speech segmentation.
\end{abstract}

\vspace{-5pt}
\section{Introduction}
\label{sec:intro}
Speaker diarization \cite{anguera2012speaker}, the task of determining ``who spoke when" in a multi-speaker audio stream has a wide range of applications from information retrieval and meeting annotations to face to face and telephonic conversation analysis. Recent speaker diarization systems \cite{garcia2017speaker, sell2018diarization} are based on segmenting the input audio stream into uniform speaker-homogeneous segments, followed by extracting fixed-length \textit{speaker embeddings} from those segments and performing speaker clustering over these embeddings.
\par
Among speaker embeddings, i-vectors \cite{shum2013unsupervised,senoussaoui2014study}, produced using generative modeling were the first employed for speaker diarization. Recently, embeddings extracted from discriminatively-trained deep neural networks (DNNs) such as d-vectors \cite{wang2018speaker, zhang2019fully}, and \textit{x-vectors} \cite{garcia2017speaker,sell2018diarization} have shown superior performance over i-vectors. These embeddings are partitioned into speaker clusters using clustering algorithms, such as Gaussian mixture models \cite{shum2013unsupervised}, mean-shift \cite{senoussaoui2014study}, agglomerative hierarchical clustering (AHC) \cite{garcia2017speaker}, k-means \cite{dimitriadis2017developing}, spectral clustering \cite{wang2018speaker, sun2019speaker} and links \cite{mansfield2018links}. All the aforementioned approaches are unsupervised in determining the number of speakers and speaker labels of a given audio session. Recently, a few supervised clustering approaches like UIS-RNN \cite{zhang2019fully} and affinity propagation \cite{yin2018neural} have also been proposed for diarization.
\par
While performances of tasks such as speech and speaker recognition have improved significantly due to supervised deep learning approaches, most of the existing diarization systems are yet to take full advantage of similar techniques. DNN-based deep clustering approaches are  popular in computer vision \cite{aljalbout2018clustering}. While appealing, they are however not immediately applied for speaker diarization tasks probably due to lack of interpretability and the problem of unknown number of speakers of a given audio session. 
Recently, deep embedded clustering on d-vectors was introduced for speaker diarization \cite{dimitriadis2019enhancements}. Incorporating the above advances, clustering with dimension reduction using non-linear neural transformation of embeddings, trained with clustering-specific loss could be beneficial for audio diarization systems. 
\par
A latent space image clustering method using generative adversarial network (GAN) along with an encoder network (\textit{ClusterGAN}) was proposed recently in \cite{mukherjee2019clustergan}. Here, the encoder network performs inverse mapping, i.e., it \textit{back-projects} the data into the latent space. 
Two main advantages of GAN-based latent space clustering are the interpretability and interpolation in the latent space \cite{mukherjee2019clustergan}. In our work, we adopt and modify this network for speaker clustering within the speaker diarization framework. The two main differences of our proposed work from \cite{mukherjee2019clustergan} are: (a) instead of random one-hot encoded variables, we use original speaker labels of the training data. Thus, the GAN generator input is a mixture of continuous random and discrete one-hot encoded speaker label variables; (b) instead of images (spectrograms), x-vector embeddings of short audio segments are used as real data input to the GAN discriminator. The GAN and encoder networks are jointly trained along with a clustering-specific loss.

\vspace{-12pt}
\section{Background}
\label{sec:format}
Over the recent years, the primary focus of research in image clustering has been to non-linearly transform the input feature space to a latent space (where the separation of data is easier) using DNNs.
Current deep clustering methods on image data include autoencoder based approaches \cite{xie2016unsupervised}, generative model based approaches such as variational deep embedding \cite{jiang2016variational} and information maximizing GAN (InfoGAN) \cite{chen2016infogan} among others. All these algorithms comprise of three essential components: deep neural network architecture, network loss, and clustering-specific loss. The network loss refers to the reconstruction loss of an autoencoder, variational loss of a variational autoencoder or the adversarial loss of GANs. It is used to learn feasible latent features and avoid trivial solutions. Clustering-specific loss can be cluster assignment losses such as k-means loss \cite{yang2017towards}, cluster assignment hardening loss \cite{xie2016unsupervised}, spectral clustering loss \cite{shaham2018spectralnet}, agglomerative clustering loss \cite{yang2016joint} or cluster regularization losses such as locality preserving loss, group sparsity loss, cluster classification loss \cite{aljalbout2018clustering}. These losses are used to learn suitable cluster-friendly representations from the data. In this work, we exploit both network loss and clustering loss in the clustering module for speaker diarization. 


\begin{figure}[!t]
    \centering
        \includegraphics[width=0.5\textwidth]{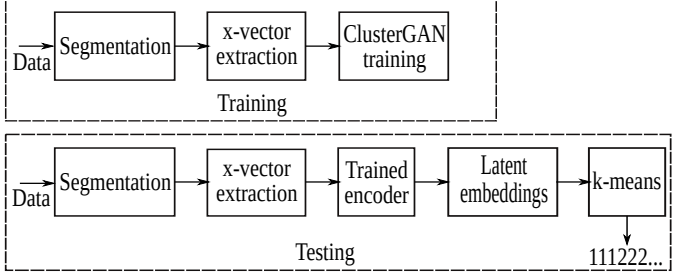}
        \caption{Schematic diagram of the proposed speaker diarization system.}\label{fig1.a}
\vspace{-10pt}
 \end{figure}  

\vspace{-10pt}
\section{Proposed speaker diarization system}
\label{sec:pagestyle}
\vspace{-5pt}
\subsection{Overview}
The overall methodology of the proposed speaker diarization system is shown in Fig. \ref{fig1.a}. The proposed system begins with the popular time-delay neural network (TDNN) speaker embedding \cite{garcia2017speaker}, i.e., x-vector extraction and followed by latent space clustering. We  discuss each module in the diarization pipeline below.

\vspace{-10pt}
\subsection{Segmentation}
Our approach starts with a temporal segmentation of 1.5 sec with 1 sec overlap. The speech segments are embedded into a fixed-dimensional x-vector of dimension 512. This TDNN-based speaker embeddings achieved state-of-the-art performance in speaker verification/diarization \cite{garcia2017speaker}. The x-vectors are then fed as inputs to the ClusterGAN network.
\vspace{-10pt}
\subsection{ClusterGAN training}
\begin{figure}[!t]
        \includegraphics[width=0.5\textwidth]{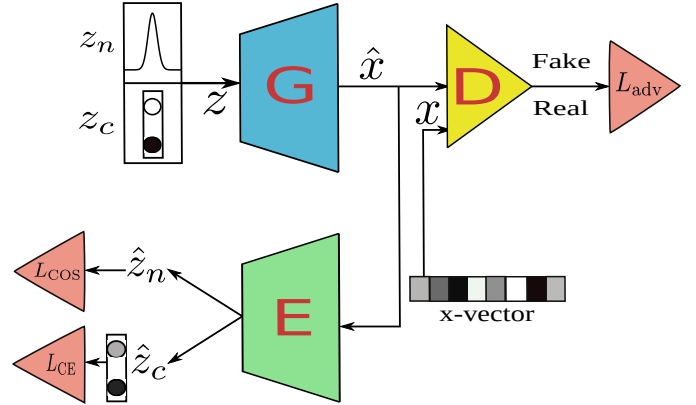}
        \caption{ClusterGAN architecture. Here, $L_{\textnormal{adv}}$, $L_{\textnormal{COS}}$ and $L_{\textnormal{CE}}$ represent adversarial, cosine distance and cross-entropy loss functions.}\label{fig1.b}
    \vspace{-12pt}
\end{figure}

The motivation behind using ClusterGAN on x-vectors is to non-linearly transform it into a lower-dimensional embedding space which is more separable. Although the idea of using a mixture of continuous and discrete latent variables as the input to GAN generator was inspired from InfoGAN \cite{chen2016infogan}, ClusterGAN is better suited for clustering than InfoGAN \cite{mukherjee2019clustergan}. ClusterGAN comprises three components: the generator ($G$), the discriminator ($D$) and the encoder ($E$), as shown in Fig. \ref{fig1.b}.
\vspace{-10pt}
\subsubsection{Adversarial training}
GANs are a recent class of deep generative models inspired by game theory metaphor, where both $G$ and $D$ networks engage in a two-player minimax game \cite{goodfellow2014generative}. The generator is considered to be a mapping from the latent space to the data space $G: z \rightarrow \hat{x}$. It takes noisy data $z$ sampled from $p_z$ and generates samples to fool the discriminator. The discriminator is considered to be a mapping from the data space to a real value $D: x \rightarrow \mathbb{R}$. It takes real data $x$ sampled from $p_x^r$ and tries to discriminate between the real and generated fake samples. We employ the improved Wasserstein GAN (IWGAN) \cite{gulrajani2017improved} for our GAN network. The objective function of this adversarial game is:
\begin{equation}
   \underset{G}{\textnormal{min}} \hspace{2pt} \underset{D} {\textnormal{max}} V_{\textnormal{IWGAN}}(D, G) = \E_{x \sim p_x^r} \left[D(x) \right] - \E_{z \sim p_z} \left[D(G(z))\right] + \lambda \cdot \textnormal{GP}
\end{equation}
where, $\lambda$ is the gradient penalty coefficient and $\textnormal{GP}$ is the gradient penalty term \cite{gulrajani2017improved}.

\vspace{-5pt}
\subsubsection{Sampling from discrete-continuous mixtures}
In order to perform clustering in the latent space, we have to back-project the data into the latent space. The latent space distribution in traditional GANs is typically chosen to be Gaussian or uniform distributions. Although such distributions contain useful information about input data distributions, they usually lead to bad clusters \cite{lipton2017precise}. To mitigate this problem, boosting the latent space using categorical variables to create non-smooth geometry is essential. However, continuity in latent space is also required for good interpolation and GANs have good interpolation ability. Therefore, we employ a mixture of continuous ($z_n$) and discrete ($z_c$) variables to the generator by concatenating $z_n$ with $z_c$. In this work, $z_n$ is randomly sampled from a normal distribution $\mathcal{N}(0, \sigma^{2} I_{d_{n}})$. We chose $\sigma = 0.1$ in all our experiments. We use the original speaker labels for the speech segments from training data as the one-hot encoded variable $z_c$. The concatenation of $z_n$ with $z_c$ enables clustering in the latent space.
\vspace{-10pt}
\subsubsection{Inverse mapping network}
Mapping from the data space to latent space is a non-trivial problem, since it requires the inversion of the generator which is a non-linear model. Existing works \cite{lipton2017precise, creswell2018inverting} tackle this problem by solving an optimization problem in $z$ to get back the latent vectors using $z^{*} = \textnormal{argmin}_{z} \mathcal{L}(G(z), x) + \lambda \lVert z \rVert _{p}$, where $\mathcal{L}$ is $L_1$ norm, $\lambda$ is a regularization constant and $\lVert \cdot \rVert _{p}$ denotes the norm. However, these approaches are not suitable for clustering since the optimization problem is non-convex \cite{creswell2018inverting, mukherjee2019clustergan}. To address this issue, an $E$ network alongside the GAN network for back-projection is introduced. We fix $z_c$ and randomly sample $z_n$ from a normal distribution with multiple restarts at each iteration step. Furthermore, to ensure precise recovery of the latent vector $z_n$, we compute the numerical difference between the encoder output latent vector $\hat{z}_n$ and $z_n$. For that, we empirically found that instead of mean square error, cosine distance is more suitable. The objective function for this task can be written as:
\begin{equation}
    \textnormal{min COS} (G, E) = \frac{1}{m} \sum_{i = 1}^m \left[1 - \frac{E(G(z_n^i)) \cdot z_n^i}{\lVert E(G(z_n^i)) \rVert \lVert z_n^i \rVert}\right]
\end{equation}
where, $m$ is the mini batch size.
\vspace{-5pt}
\subsubsection{Clustering-specific loss}
To learn cluster friendly representations, we incorporate cluster classification loss while training as cross-entropy (CE) loss. The soft-max layer output obtained by $E$ network is used for computing the cross-entropy loss. This loss encourages the latent embeddings to cluster and hence increase the discriminative information. We minimize this cross-entropy loss as:
\begin{equation}
   \textnormal{min CE} (G, E) = \frac{1}{m} \sum_{i = 1}^m \left[p(z_{c, i}^k) \textnormal{log} \hspace{2pt} p\left(E(G(z_{c, i}^k))\right)\right] 
\end{equation}
where, the first term is the empirical probability that the embedding belongs to the $k$-th speaker, and the second term is the predicted probability (by the encoder) that the embedding belongs to the $k$-th speaker.
\vspace{-5pt}
\subsubsection{Joint training}
We train the GAN and encoder networks jointly. The training objective function takes the following form:
\begin{equation}
    \underset{G, E}{\textnormal{min}} \hspace{2pt} \underset{D} {\textnormal{max}} \left[a \cdot V_{\textnormal{IWGAN}}(D, G) + b \cdot \textnormal{COS} (G, E) + \\ c \cdot \textnormal{CE} (G, E) \right]
\end{equation}
The weights $b$ and $c$ are used to control the importance of preserving continuous and discrete latent variables. Algorithm \ref{algo1} shows the training steps of ClusterGAN.

\begin{algorithm}[!t]
\caption{ClusterGAN algorithm. Default values: $\lambda$ = 10, $m$ = 64, $n_{\textnormal{critic}}$ = 5, $\alpha$ = $1\mathrm{e}{-4}$, $\beta_1$ = 0.5, $\beta_2$ = 0.9}
\label{algo1}
\begin{algorithmic}[1]
\Require{$\lambda$: gradient penalty coefficient; $\alpha$: learning rate; $m$: batch size; $N_{\textnormal{it}}$: number of iterations; $n_{\textnormal{critic}}$: number of critic iterations for each generator iteration; $\alpha$, $\beta_1$, $\beta_2$: Adam hyper-parameters} {}

\For{$it$ = 1 to $N_{\textnormal{it}}$}
\For{$\tau$ = 1 to $n_{\textnormal{critic}}$}
\State Sample $\{x^{(i)}\}_{i = 1}^m$, a batch of x-vectors
\State Update the discriminator parameters by 
\State \vspace{-20pt} \begin{multline*} \hspace{17pt}\theta \leftarrow \textnormal{Adam}[\triangledown_{\theta} \{\frac{1}{m}\sum_{i = 1}^m a \cdot [D_{\theta}(x) - D_{\theta}(G_{\phi}(z)) + \\ \vspace{-10pt}\lambda \cdot \textnormal{GP}]\}, \theta, \alpha, \beta_1, \beta_2]\end{multline*} 
\vspace{-20pt}
\EndFor
\State Sample $\{z^{(i)}\}_{i = 1}^m$, a batch of latent vectors
\State Update the generator and encoder parameters by 
\State \vspace{-20pt} \begin{multline*}  \vspace{-10pt}\hspace{5pt} \phi, \psi \leftarrow \textnormal{Adam}[\triangledown_{\phi, \psi} \{ \frac{1}{m}\sum_{i = 1}^m - a \cdot D_{\theta}(G_{\phi}(z^{(i)})) +  \\ b \cdot \textnormal{COS} (G_{\phi}, E_{\psi}) + c \cdot \textnormal{CE} (G_{\phi}, E_{\psi})\}, \phi, \psi, \alpha, \beta_1, \beta_2 ] \end{multline*} 
\vspace{-20pt}
\EndFor
\end{algorithmic}
\end{algorithm}

\subsection{Testing}
After offline training, only the trained encoder model is required to produce the proposed latent embeddings for the input x-vectors of a test audio session. The concatenated latent embeddings ($z_n$ and $z_c$) are then clustered to produce speaker labels of each segment using k-means. 

\vspace{-10pt}
\section{Experimental evaluation}
\label{sec:typestyle}
\subsection{Data preparation}
\vspace{-5pt}
We evaluate our proposed algorithm on the AMI meeting corpus and two child-clinician interaction corpora: ADOS \cite{bone2012spontaneous} and BOSCC \cite{kumar2018knowledge}. The AMI database consists of 171 meetings recorded at four different sites (Edinburgh, Idiap, TNO, Brno). For our evaluation, we use the official speech recognition partition of AMI dataset\footnote{\url{http://groups.inf.ed.ac.uk/ami/download/}}. We exclude the TNO meetings from dev and eval set, which is a common practice in diarization studies \cite{sun2019speaker, yella2015comparison}. The details of the dataset partition are shown in Table \ref{ami:data}.
\par
The ADOS \cite{lord2000autism} is a diagnostic tool which comprises over 10 play-based, conversational tasks. We chose two conversational tasks: \textit{Emotions} and \textit{Social Difficulties and Annoyance} from 272 sessions for our evaluation. BOSCC \cite{grzadzinski2016measuring} is a new treatment outcome measure, also comprised of play-based, conversational segments. For this study, 24 BOSCC sessions are selected.



\begin{table}[!t]
\vspace{-10pt}
\caption{Details of the AMI data set used for our experiments.}
\vspace{-10pt}
\centering
\setlength{\tabcolsep}{10pt}
\begin{tabular}{c|cc}
\Xhline{2.5\arrayrulewidth}
      & \#Meetings & \#Speakers          \\ \Xhline{2.5\arrayrulewidth}
Train & 136     & 155                 \\
Dev   & 14      & 17  \\
Eval  & 12      & 12  \\ \Xhline{2.5\arrayrulewidth}
\end{tabular}\label{ami:data}
\vspace{-15pt}
\end{table}

\vspace{-5pt}
\subsection{Experimental framework}
\vspace{-2pt}
\subsubsection{Baseline systems}
Since our proposed system uses x-vectors as input features, we used the Kaldi-based AHC clustering with PLDA scoring on x-vectors \cite{garcia2017speaker} (denoted as x-vector in this paper) as our main baseline system. We also show results on x-vectors with k-means clustering (denoted as k-means in this paper), as our second baseline.
\vspace{-10pt}
\subsubsection{Model specifications}
In all our systems, x-vectors are extracted using the Voxceleb\footnote{\url{https://kaldi-asr.org/models/m7}} models available in the Kaldi recipe. Diarization performance of the proposed system is evaluated for two models trained with different amounts of supervised data: P1 and P2. P1 is trained only on the AMI train set, whereas P2 is trained on both AMI train set and 60 beamformed ICSI \cite{janin2003icsi} sessions with a total number of 46 speakers. The generator and discriminator networks in the proposed systems are simple feed forward neural networks with one and two hidden layers respectively each with 512 nodes. 
The input layer of $G$ consists of $d = d_n + d_c$ nodes ($d_n$, $d_c$ are the dimensions of $z_n$ and $z_c$ respectively), where $d_n = 30$ for both P1 and P2 models, and $d_c = 155$ for P1 and $201$ for P2 model.
$G$'s output layer has 512 nodes, which is the x-vector dimension. The input and output layer of $D$ contains 512 nodes and one node, respectively. On the other hand, the $E$ network consists of a single hidden layer with 512 nodes and input layer is linear with 512 nodes. 
The output layer of $E$ is a linear layer with $d$ nodes from which the first $d_n$ nodes are directly used as $\hat{z}_n$ and the rest are passed through a soft-max layer to produce $\hat{z}_c$.
For all the three networks, the activation function in the hidden layers is ReLU. In the proposed system, we use the original speaker labels from the training data to produce $z_c$ for each segment. The networks are optimized using Adam \cite{kingma2014adam} with a mini-batch size of 64 samples and learning rate $1\mathrm{e}{-4}$. We fixed the $a$, $b$ and $c$ values as 1, 2 and 10 respectively by tuning on AMI dev set. The number of iterations is fixed to 30k based on optimizing DER on the AMI dev set.

\begin{figure*}[!t]
 \centering
  \includegraphics[width=\textwidth]{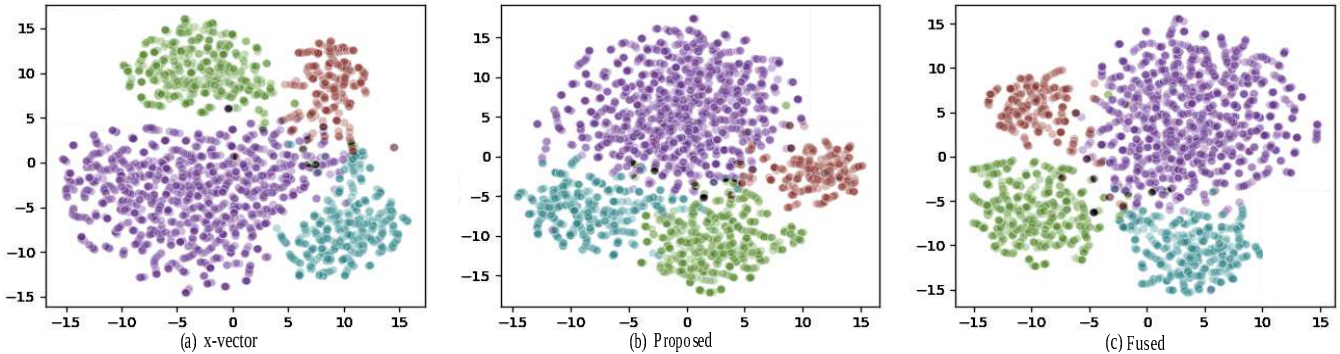}
  \vspace*{-0.6cm}
  \caption{TSNE visualization of (a) x-vector, (b) proposed and (c) fused embeddings of IS1008a AMI session. This AMI session contains four speakers and each speaker is represented by different colours in the figure.}
  \label{fig:tsne}
  \vspace{-12pt}
\end{figure*}

\vspace{-10pt}
\subsubsection{Performance metrics}
The performance of speaker diarization systems is evaluated by using NIST diarization error rate (DER) \cite{fiscus2006rich}, which is typically calculated with a 0.25 sec collar. Since the primary focus of this paper is on the effectiveness of new speaker embeddings in clustering, likewise in \cite{yella2015comparison, garcia2017speaker, sun2019speaker}, for all the experiments in this paper we use oracle speech activity detection (SAD). Therefore, all DER values reported in this work correspond to speaker confusion errors with no missed or false alarm speech. 

\vspace{-10pt}
\subsection{Results and discussions}

\begin{table}[!t]
\caption{Results on AMI dev and eval set for the baseline and proposed systems.}
\vspace{-10pt}
\centering
\setlength{\tabcolsep}{8pt}
\begin{tabular}{ccccc}
\Xhline{2.5\arrayrulewidth}
\multirow{2}{*}{System} & \multicolumn{2}{c}{\begin{tabular}[c]{@{}c@{}}Avg. DER (in \%) \\ (oracle SAD, \\ known \#speakers)\end{tabular}} & \multicolumn{2}{c}{\begin{tabular}[c]{@{}c@{}}Avg. DER (in \%) \\ (oracle SAD, \\ estimated \#speakers)\end{tabular}} \\
                        \cline{2-3} \cline{4-5} & Dev                              & Eval                             & Dev                                & Eval                               \\ \Xhline{2.5\arrayrulewidth}
x-vector                & 11.65                            & 11.34                            & 11.08                              & 10.37                              \\
k-means                 & 11.94                            & 11.45                            & 12.64                                     & 12.26                                    \\
P1                      & 10.17                            & 10.10                            &   10.98                                 &    11.26                                \\
P2                      & 9.67                             & 11.64                            &   10.33                                 &      11.56                              \\
x-vector + P1           & 7.45                             & \textbf{7.82}                             & 8.73                                   & 9.11                                   \\
x-vector + P2           & \textbf{6.98}                            & 8.85                             & \textbf{7.93}                                   &  \textbf{8.92}                                  \\
Sun et. al. \cite{sun2019speaker}      & --                               & --                               & 12.22                              & 12.99       \\
\Xhline{2.5\arrayrulewidth}
\end{tabular}\label{table1}
\vspace{-15pt}
\end{table}

\par
\vspace{-5pt}
\subsubsection{Results on AMI dev and eval set}
\vspace{-5pt}

Results for diarization performance on AMI dev and eval sets are reported in Table \ref{table1}. We show results for oracle SAD with both known number of speakers and estimated number of speakers. For the x-vector baseline, we use thresholding on the PLDA scores to perform AHC clustering for unknown number of speakers. The number of speakers for k-means and proposed systems are estimated using Eigen-gap analysis of the affinity matrix constructed from the cosine distance of x-vector embeddings followed by binarization and symmetrization \cite{park2019second}. From Table \ref{table1} column 2, we see that for known number of speakers, the P1 system beats x-vector (state-of-the-art) and k-means systems for both AMI dev and eval sets. The performance improves further after incorporating fusion with x-vector embeddings ((x-vector + P1) and (x-vector + P2)). 
It is observed that both the fused systems significantly outperform all the other systems. The best achieved DER using our fused systems on AMI dev and eval set are 6.98\% and 7.82\% respectively. This is attributed to the fact that our proposed embeddings have complementary information with x-vector embeddings. 

We report the diarization performance of all the systems for estimated number of speakers in Table \ref{table1} column 3. Surprisingly, it is observed that x-vector baseline system with thresholding on the PLDA scores for AHC clustering produces a slightly better performance as compared to the oracle number of speaker condition. 
In contrast, all the other methods' performance degrades for estimated number of speakers. We also compare the proposed diarization system with the work proposed in Sun et al. \cite{sun2019speaker} evaluated on the same data set. The system proposed in \cite{sun2019speaker} is a 2D self-attentive combination of d-vectors with spectral clustering back-end. As seen in Table \ref{table1} column 3, our proposed and x-vector fused embeddings with k-means clustering back-end outperforms other baseline methods.

\begin{table}[!t]
\caption{Results on ADOS and BOSCC databases for the baseline and proposed systems.}
\vspace{-10pt}
\centering
\setlength{\tabcolsep}{10pt}
\begin{tabular}{ccc}
\Xhline{2.5\arrayrulewidth}
System        & \begin{tabular}[c]{@{}c@{}}Avg. DER (in \%) \\ on ADOS\end{tabular} & \begin{tabular}[c]{@{}c@{}}Avg. DER (in \%) \\ on BOSCC\end{tabular} \\
\Xhline{2.5\arrayrulewidth}
x-vector      & 14.36                         & 21.69                            \\
k-means       & 12.35                         & 14.73                            \\
P1            & 11.27                         & 14.63                            \\
P2            & 11.08                         & 13.35                            \\
x-vector + P1 & 9.38                          & 13.55                            \\
x-vector + P2 & \textbf{9.22}                          & \textbf{11.17}                           \\
\Xhline{2.5\arrayrulewidth}
\end{tabular}\label{table2}
\vspace{-15pt}
\end{table}

\vspace{-5pt}
\subsubsection{TSNE visualization}
\vspace{-5pt}
We show TSNE visualizations of x-vector, proposed and fused embeddings of AMI session IS1008a in Fig. \ref{fig:tsne}. It is evident from the figure that the proposed embedding based clusters are slightly more compact as compared to the x-vectors. However, fused embedding based clusters are the most compact within a class and most separated between classes.

\vspace{-5pt}
\subsubsection{Generalization ability}
\vspace{-5pt}
From Table \ref{table2}, we observe significant performance improvement for the proposed system over the baselines on both ADOS and BOSCC sessions. In addition, the P2 model which is trained on more data achieves better performance than P1 for both individual and fused scenarios. In particular, the improvement is notable compared to the x-vector baseline. 
We hypothesize that the PLDA model pre-trained on Voxceleb presents a significant domain mismatch in this case. Moreover, both P1 and P2 systems, either used individually or in fusion with x-vectors, are superior to k-means. The best system (x-vector + P2) achieves a relative 36\% and 49\% improvement over x-vector on those two databases.
\vspace{-10pt}
\section{Conclusions}
\vspace{-5pt}
We presented a new deep latent space clustering using ClusterGANs to perform speaker diarization. The entire system was trained in a supervised manner along with a clustering-specific loss function. We observed that ClusterGAN-based latent embeddings provide superior performance than x-vector embeddings. Further improvement was achieved after fusing proposed and x-vector embeddings. Experimental results showed a significant DER reduction for the proposed system over state-of-the-art x-vector diarization system on AMI, ADOS and BOSCC corpora. Future work could use spectrograms directly instead of pre-trained embeddings as the GAN discriminator input. 
\vfill\pagebreak

\bibliographystyle{IEEEbib}
{\bibliography{refs}}

\end{document}